\documentclass[
 reprint,
 amsmath,amssymb,
 aps,
 prd,
]{revtex4-2}

\usepackage{graphicx}% Include figure files
\usepackage{dcolumn}% Align table columns on decimal point
\usepackage{bm}% bold math
\usepackage{hyperref}% add hypertext capabilities
\usepackage{aas_macros}
\usepackage{xcolor}
\usepackage{gensymb}
\graphicspath{{./}{figures/}}

\begin{document}

\title{The proper way to spatially decompose the gravitational-wave origin in stellar collapse simulations}%

\author{Shuai Zha}\email{zhashuai@ynao.ac.cn}
\affiliation{Yunnan Observatories, Chinese Academy of
Sciences (CAS), Kunming 650216, China}
\affiliation{Key Laboratory for the
Structure and Evolution of Celestial Objects, CAS, Kunming 650216,
China}
\affiliation{International Centre of Supernovae, Yunnan Key Laboratory,
Kunming 650216, China}

\date{\today}% It is always \today, today,

\begin{abstract}
Gravitational waves (GWs) hold great potential for an unobscured view of protoneutron stars (PNSs) formed as a result of stellar collapses. While waiting for discovery, deepening the understanding of GW emission in theory is beneficial for both optimizing searching strategies and deciphering the eventual data. One significant aspect is the spatially dependent contribution to the overall GW signal extracted from sophisticated hydrodynamic simulations. I present the proper way to perform the spatial decomposition of GW strain with the quadrupole formula in the slow-motion and weak-field approximation. Then I demonstrate the approach using the results of a 2D axisymmetric pseudo-Newtonian hydrodynamic simulation of core-collapse supernova. I show a detailed comparison between the proper and improper methods and discuss the possible consequences based on the improper method. Moreover, with the correct approach, the GW spatial profiles agree well with those calculated from a consistent perturbative method. 
\end{abstract}

\maketitle

\section{Introduction}
Since 2015 we have formally entered the era of gravitational-wave (GW) astronomy with the groundbreaking detection of GWs from the coalescence of binary black holes, i.e. GW150914 \cite{gw150914}. The field flourished with the joint observation of the $\gamma$-ray burst, kilonova, and GWs from a binary neutron-star merger event \cite{gw170817,gw170817gamma,gw170817mma}, a victory of multi-messenger astronomy. More recently, several international teams reported evidence of nano Hz GW background using pulsar timing arrays \cite{CPTA,EPTA,NANOgrav,PPTA}, with important implications for cosmology and merging supermassive black holes. Yet another historic and frequently discussed candidate GW source, i.e. collapse and explosion of stars, is still waiting for discovery \cite{gossan2016,ligo_sn}. To our latest understanding, current ground-based detectors \cite{ligo,virgo,kagra} can capture GWs from stellar collapse events in the Milky Way (e.g. \cite{abdikamalov22,mezzacappa2024}). Due to the low galactic supernova rate ($\sim$2-3 per century \cite{adams2013}), it may be a long time for such GW observations to come. We look forward to the further development of next-generation detectors desperately for more distant targets, such as the Neutron-star Extreme Matter Observatory (NEMO, \cite{NEMO}), the Cosmic Explorer (CE, \cite{CE}) and the Einstein Telescope (ET, \cite{ET}).

\begin{figure*}
    \centering
    \includegraphics[width=0.45\textwidth]{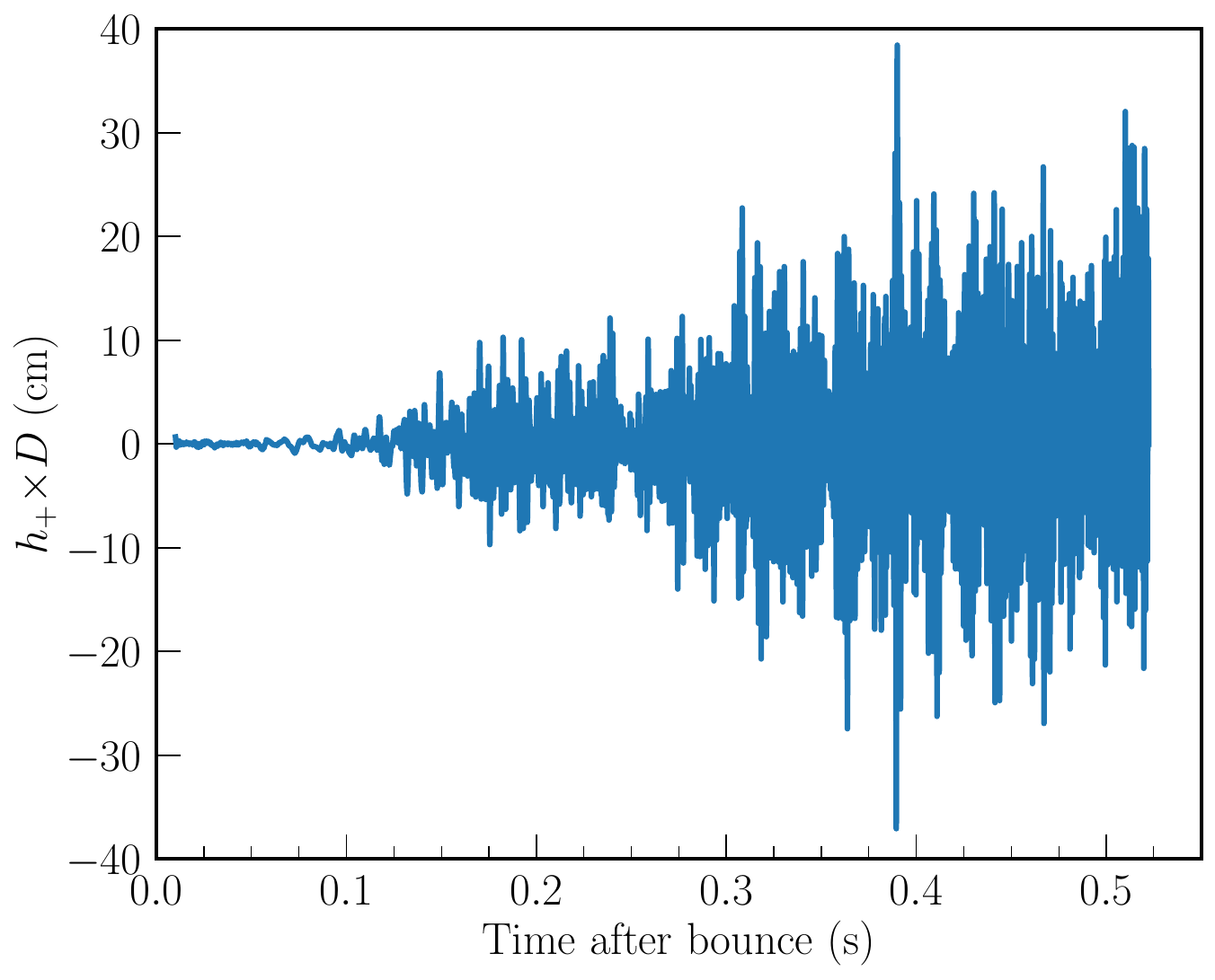} \hspace{0.05\textwidth}
    \includegraphics[width=0.45\textwidth]{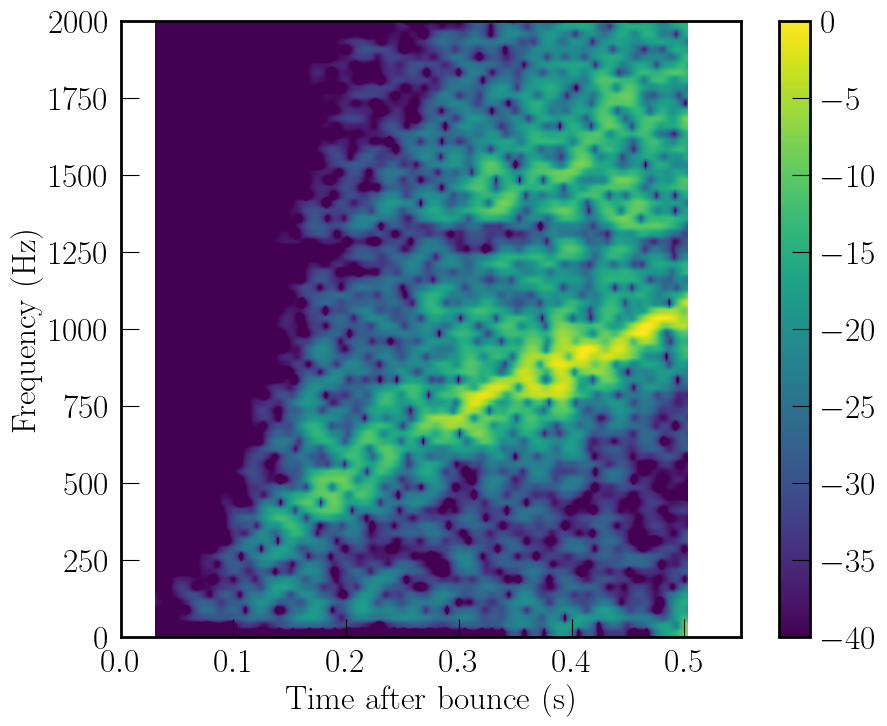}
    \caption{The gravitational-wave (GW) signal results from a 2D axisymmetric simulation with a 20\,$M_{\odot}$ progenitor \cite{woosley07} and SFHo equation of state \cite{sfho}. Left panel: the time-domain waveform, i.e. GW strain amplitude as a function of time. Right panel: the GW power spectrogram which results from the short-time Fourier transform of the time-domain signal with a moving temporal window of 40 ms by the Python function `matplotlib.pyplot.specgram'. Note that the contribution of the prompt convection is suppressed during the first 100\,ms after bounce because the simulation first runs in 1D spherical symmetry and switches to 2D axisymmetry at $\sim10$\,ms after bounce.}
    \label{fig:gw}
\end{figure*}

For now, understanding the GW characteristics of stellar collapses through theoretical investigations is of vital importance to prepare for the eventual discovery. The collapse of stars involves all 4 known fundamental interactions and is one of the most challenging mysteries in modern physics and astronomy \cite{bethe1990}. Lacking thermal support from nuclear fusions, the stellar core contracts dynamically until the formation of a protoneutron star (PNS), in which strong forces and nuclear degeneracy balance the gravity \cite{oertel2017}. Quasi-normal quadrupolar ($l=2$) oscillations of the PNS can lead to GW emission with a frequency ranging from hundreds to thousands of Hz \cite{sotani17,morozova18,torres18}. Fig.~\ref{fig:gw} shows an example of such GW signals extracted from a 2D axisymmetric simulation, with the time-domain waveform (left panel) and power spectrogram (right panel). The amplitude of GW strain evolves stochastically with time while the GW frequency ramps up as a result of the PNS contraction. GW amplitudes in 3D simulations can be 10 times lower than those in 2D simulations with axisymmetry while the frequency evolution seems to be far less dependent on dimensionality \cite{andresen17,radice19,mezzacappa23} and resolution \cite{morozova18}.

It is yet unclear what excites the PNS oscillations in relation to the peak GW emission \cite{murphy09,muller13,andresen17,radice19,mezzacappa23}. There are two major compelling proposals that the excitation comes from the downflows in the gain region impinging the PNS \cite{murphy09,muller13,radice19} or from the PNS convective motions \cite{andresen17,mezzacappa23}. Evidence for the first mechanism is that the peak GW frequency coincides with the downflow timescales \cite{murphy09} and the Brunt-V\"ais\"al\"a frequency of the convectively stable layer just above the PNS convective zone \cite{muller13}. It has further support from the proportion between the radiated GW energy and the total turbulent energy accreted by the PNS \cite{radice19}. On the other hand, the second mechanism is built on that the dominant contribution to GWs comes from the PNS convective zone \cite{andresen17,mezzacappa20}. The convective overshooting layer can act as frequency stabilizer \cite{andresen17}  or an additional exciter \cite{mezzacappa23} by jointly analyzing the complex fluid motions and the spatial contribution of GW emission. However, there is a caveat for their analyses of the spatial decomposition of the GW emission as pointed out briefly in the appendix of Ref.~\cite{oliver21}. Ref.~\cite{mezzacappa23} followed a similar approach and compared the proper and improper methods. They found that the caveat is not a crucial issue by comparing decomposed time-domain waveforms and utilized the original approach for the main discussion. Here, this work revisits this caveat and demonstrates the proper approach with numerical experiments, especially a detailed frequency-domain analysis.

Another way to understand the GWs from stellar collapse simulations is by solving hydrodynamic equations in the perturbative regime, dubbed supernova seismology \cite{fuller15}. Mode frequency matching illustrated a few gravity modes and the fundamental mode responsible for the peak GW emission, though the nature of these modes is under debate \cite{sotani17,morozova18,torres18}. Moreover, direct comparisons of radial profiles between simulations and perturbative analyses for the PNS oscillations indicated a global GW emission picture in stellar collapses \cite{ryan2020,zha24}. As will be clear in this work, this comparison is only possible with the proper spatial decomposition of the GW emission within simulation data.

Overall, this paper complements Zha et al. (2024) \cite{zha24} with the proper formula to spatially decompose the GW emission and demonstrate its usage in stellar collapse simulations. The paper is organized as follows. I present the formulae for the spatial decomposition of GW emission in \S\,\ref{sec:form}. I analyze and present the numerical results in \S\,\ref{sec:res}. I conclude my findings and give an outlook in \S\,\ref{sec:conclu}.

\section{Formulae} \label{sec:form}
In this section, I present the formulae for decomposing the spatial contribution to the GW emission, or more specifically the GW strain contributed by successive spherical shells, in 2D axisymmetric stellar collapse simulations. This is relevant to the analysis in \S\,\ref{sec:res}. One can derive the corresponding formulae in 3D simulations accordingly (see, e.g.,  \cite{mezzacappa23}) which I leave for future studies due to the unaffordable amount of required computational resources.

It is a formidable task to simulate stellar collapses in the fully general relativistic framework together with multiple dimensions, finite-temperature nuclear equation of state, and spectral neutrino transport \cite{muller2020}. Instead, simulations are usually performed in Newtonian hydrodynamics with a pseudo relativistic gravity \cite{2006A&A...445..273M,muller08} or general-relativistic hydrodynamics with the conformally flat approximation \cite{banyuls1997,dimmelmeier02,pablo05,muller13}. Then, one can extract GW strain from such numerical simulations with the quadrupole formula within the weak-field and slow-motion approximations \cite{finn90,moenchmeyer91}. Here, I work with the Newtonian hydrodynamic variables and equations and one can derive the general relativistic form in the same way according to, e.g., \cite{dimmelmeier02,muller13}.  

In 2D axisymmetry, the only non-vanishing term is
\begin{equation}
    h_+ = \dfrac{3}{2} \dfrac{G}{Dc^4} \sin^2{\Theta} \dfrac{\mathrm{d}^2}{\mathrm{d}t^2} I_{zz},
\end{equation}
where $\Theta$ is the angle between the line of sight and the symmetry axis ($z$), and I set $\Theta=90^{\degree}$ for simplicity. The subscript $+$ denotes the plus polarization. $I_{zz}$ is the trace-free quadrupole moment:
\begin{equation} \label{eq:Izz}
    I_{zz} = \dfrac{2}{3} \int \rho r^2 \mathcal{P}_2(\cos\theta) \mathrm{d}V,
\end{equation}
where $r$ is the spherical radius, $\theta$ is the angle between the positive $z$-axis and line segment, $\mathcal{P}_2$ is the Legendre polynomial of degree 2, and $\mathrm{d}V$ is the volume element. To get the overall GW emission, one performs the integration over the whole star in Eq.~(\ref{eq:Izz}). In practice, one can use a finite outer radius in the integration which is large enough for the convergence of GW strain and minimizes numerical noises.

I denote the approach that takes numerical differentiation of $I_{zz}$ twice directly as the method QF2. This may introduce high-frequency noises due to finite differences, especially for tiny time steps \cite{finn90}. Alternatively, one can reduce the numerical differentiation with the help of the mass conservation equation
\begin{equation}
     \dfrac{\partial \rho}{\partial t} + \nabla \cdot (\rho \vec{v}) = 0,
\end{equation}
so that 
\begin{equation} \label{eq:dIdt}
    \dfrac{\mathrm{d}I_{zz}}{\mathrm{d}t} = \frac{2}{3} \int \rho r \Big(2v_r\mathcal{P}_2(\cos\theta) +v_{\theta}\dfrac{\mathrm{d}\mathcal{P}_2(\cos\theta)}{\mathrm{d}\theta}\Big) \mathrm{d}V,
\end{equation}
where $v_r$ and $v_{\theta}$ are the velocities along $r$- and $\theta$-directions, respectively. I denote the approach that takes numerical differentiation of $\mathrm{d}I_{zz}/\mathrm{d}t$ to get $h_+$ as the method QF1. 

A common miss is that Eq.~(\ref{eq:dIdt}) results from partial integration and there are additional terms at the upper and lower limit of the integral. For a definite integral inside a layer from $r_1$ to $r_2$, Eq.~(\ref{eq:dIdt}) reads:
\begin{equation} \label{eq:dIdt_layer}
\begin{aligned}
    \dfrac{\mathrm{d}I_{zz}}{\mathrm{d}t}\Big|_{r_1\to r_2} &= -\frac{2}{3}\int \Big(r^4\rho v_r|_{r_1}^{r_2} \Big)\mathcal{P}_2(\cos\theta)\mathrm{d}\Omega \\ 
    &+\frac{2}{3} \int_{r_1}^{r_2} \rho r \Big(2v_r\mathcal{P}_2(\cos\theta) +v_{\theta}\dfrac{\mathrm{d}\mathcal{P}_2(\cos\theta)}{\mathrm{d}\theta}\Big) \mathrm{d}V,
\end{aligned}
\end{equation}
where $\mathrm{d}\Omega$ is the solid angle element, so $\mathrm{d}V=r^2 \mathrm{d}r\mathrm{d}\Omega$. I denote the first part in Eq.~(\ref{eq:dIdt_layer}) as the surface term and denote the approach QF1 including this surface term as QF1$^{*}$. 

If one takes $r_1$ and $r_2$ equal to 0 and infinity, the surface term vanishes as $\rho=0$ at infinity. This means that QF1 and QF1$^*$ are equivalent for evaluating the overall GW emission. However, when decomposing the overall GW emission into the contribution of successive layers as done in \cite{andresen17,mezzacappa20,mezzacappa23}, only QF2 and QF1$^{*}$ are mathematically valid while QF1 will mistakenly attribute the spatial contribution. Note that the surface term in Eq.~(\ref{eq:dIdt_layer}) is from rigorous mathematical derivation. One should not confuse it with determining the boundaries of successive layers, such as the PNS convective and overshooting regions.

\section{Numerical results} \label{sec:res}
\subsection{Stellar collapse simulation} \label{ssec:sim}
\begin{figure*}
    \centering
    \includegraphics[width=0.46\textwidth]{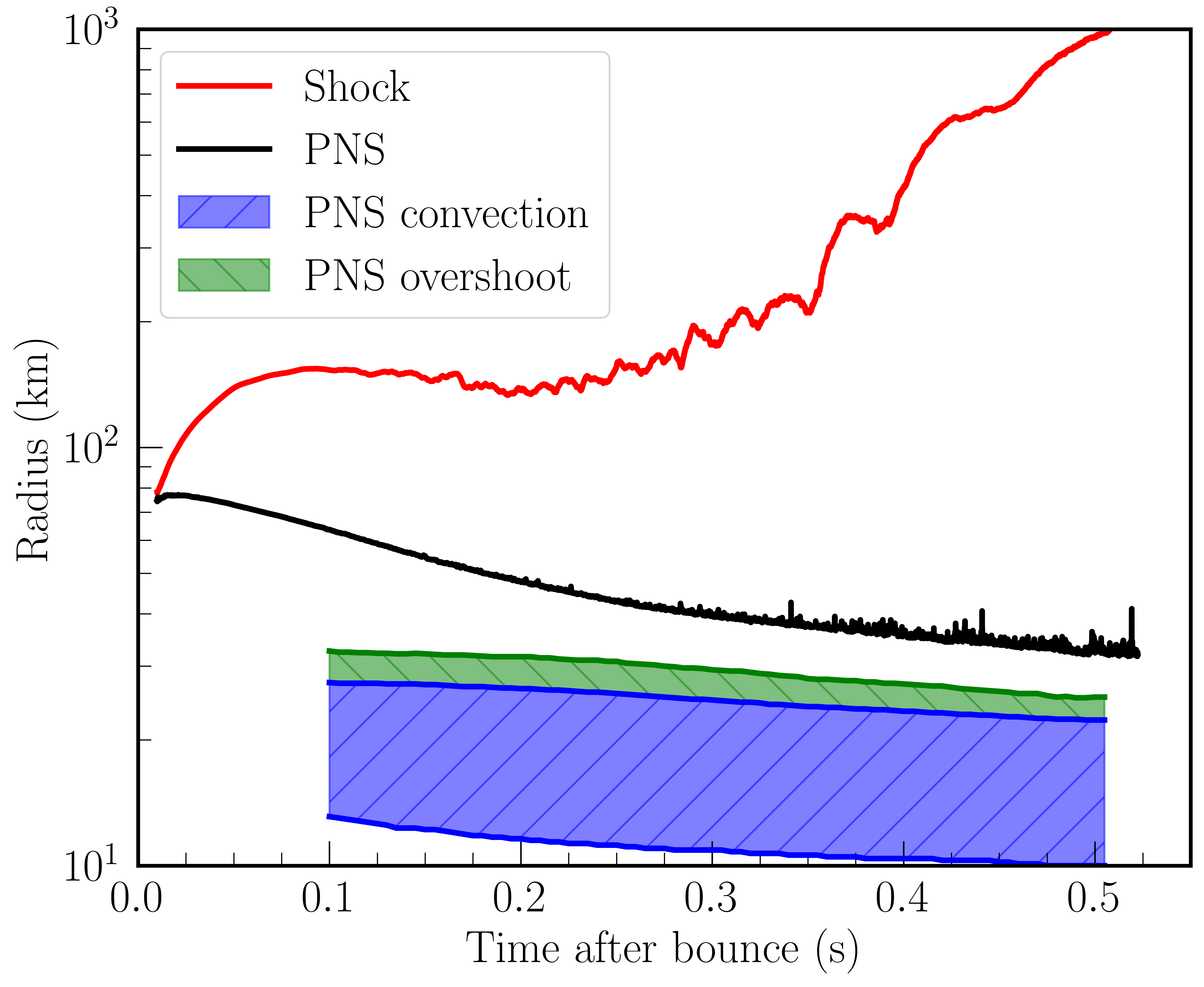}
    \hspace{0.02\textwidth}
    \includegraphics[width=0.475\textwidth]{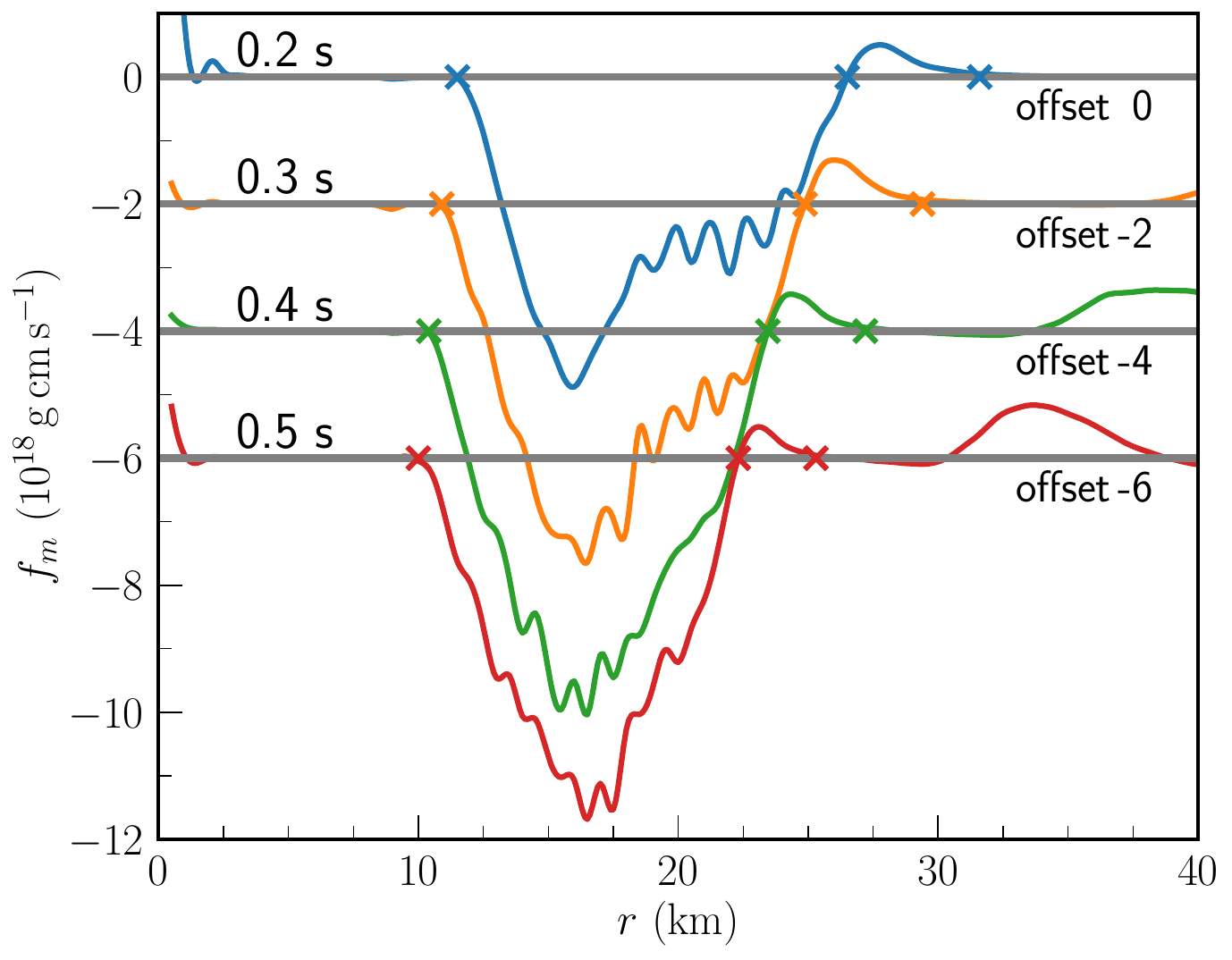}
    \caption{Left panel: Time evolution of several important radii. Shock stands for the mean shock radius, PNS stands for the protoneutron star radius defined as the locus where the spherically averaged density equals 10$^{11}$\,g\,cm$^{-3}$. PNS convection and overshoot denote the PNS convective and overshooting regions, respectively. Right panel: Turbulent mass fluxes (Eq.~(\ref{eq:fm})) at four time points with offsets for a clear presentation. The fluxes are averaged over a 40 ms window to reduce numerical noises. Crosses ($\times$) with the same color as lines mark the boundaries of the PNS convective and overshooting regions, as defined in the main text.}
    \label{fig:sim}
\end{figure*}

To demonstrate the approach of \S\,\ref{sec:form}, I ran a 2D axisymmetric stellar collapse simulation with FLASH v.4 \cite{FLASH}. The hydrodynamic equation is Newtonian supplemented by a multipole ($l\mathord{=}16$) gravitational potential \cite{couch13} whose monopole term is modified by the Case A formula for relativistic approximation \cite{2006A&A...445..273M}. The simulation employs a non-rotating solar-metallicity 20\,$M_\odot$ progenitor model \cite{woosley07} and the SFHo equation of state \cite{sfho}. Neutrino transport with full velocity dependence is solved using a 3-species ($\nu_e,~\bar{\nu}_e,~\nu_x\mathord{=}\{\nu_\mu,\bar{\nu}_\mu,\nu_\tau,\bar{\nu}_\tau\}$) two-moment scheme with the `M1' closure \cite{oconnor18a}. The transport uses 12 groups for neutrino energy ranging from 0 to 250\,MeV logarithmically with their opacities generated by the \texttt{NuLib} library \cite{gr1d} following the interactions and prescriptions considered in Ref.~\cite{2018JPhG...45j4001O}. The simulation runs in 1D spherical symmetry for the collapse phase and switches to 2D axisymmetry at 10\,ms after bounce. The simulation terminates at $\sim0.53$\,s postbounce and I record detailed profiles of density, velocities, etc. every 20\,$\mu$s for the purpose of spatially decomposing the GW emission. The 2D simulation grid uses the cylindrical coordinate system, covering a box of $10^4$\,km and $\pm10^4$\,km in the cylindrical $r$ and $z$ direction, respectively. With adaptive mesh refinement, the spatial resolution is $\sim250$\,m inside $\sim$80\,km to resolve the PNS region well and maintains an effective angular resolution of at least $\sim0.6^{\degree}$ outwards.

The evolution of the mean shock radius ($R_{\rm sh}$) and PNS radius ($R_{\rm PNS}$) is rendered in the left panel of Fig.~\ref{fig:sim}. The simulation terminates when $R_{\rm sh}$ expands rapidly and exceeds 1000\,km that indicates a successful explosion. $R_{\rm PNS}$ is defined as the locus with the spherically averaged density equal to $10^{11}\,{\rm g~cm^{-3}}$.  $R_{\rm PNS}$ decreases from a maximum value of $\sim75$\,km at $\sim0.02$\,s postbounce to $\sim30$\,km at $\sim0.5$\,s postbounce due to joint effects of neutrino cooling and mass accretion. 

The left panel of Fig.~\ref{fig:sim} also shows that below $R_{\rm PNS}$, there is a convectively unstable region (blue-shaded area) due to negative gradients in the radial profiles of specific entropy and lepton number fraction \cite{dessart06,marek09}, with an overshooting region (green-shaded area) on top of it. I omit the epoch before 0.1\,s postbounce when PNS convection has not fully developed. Previous studies \cite{andresen17,mezzacappa20,mezzacappa23} decomposed the overall GW emission to the contribution of these regions. To implement this conventional analysis, I determine the boundaries of the convective and overshooting regions similarly to Andresen et al. \cite{andresen17}. The definition of volume-weighted horizontal averages of any hydrodynamic variable $X$ is:
\begin{equation}
    \langle X \rangle = \dfrac{\int X \mathrm{d}\Omega}{\int  \mathrm{d}\Omega}.
\end{equation}
Angular fluctuation of $X$ at a fixed radius is then:
\begin{equation}
    X' = X - \langle X \rangle.
\end{equation}
With this definition, I calculate the turbulent mass flux $f_m$ as follows:
\begin{equation} \label{eq:fm}
    f_m = \langle \rho^\prime v_r^\prime. \rangle
\end{equation}
Convective regions have negative fluxes while overshooting regions have positive fluxes as the incoming inertial material is denser than the surroundings. The right panel of Fig.~\ref{fig:sim} shows the turbulent mass fluxes calculated at 0.2, 0.3, 0.4 and 0.5\,s postbounce, averaged over a 40\,ms window to reduce numerical noises. The crosses mark the boundaries of the PNS convective and overshooting regions, determined as follows. The lower boundary of the convective region (left cross) has $f_{m} = 0.01 \times f_{m,\min}$, where $f_{m,\min}$ is the minimum turbulent mass flux. Its upper boundary (middle cross, also the lower boundary of the overshooting regions) has $f_{m} = 0$ which results from a cubic spline interpolation. The upper boundary of the overshooting region (right cross) has $f_{m} = 0.1 \times f_{m,\max}$, where $f_{m,\max}$ is the maximum turbulent flux in the overshooting region. Note that these definitions are not unambiguous and here I follow the approach of Andresen et al. \cite{andresen17}. This uncertainty is not important in the global emission scenario as will be presented in \S\,\ref{ssec:global}.

Fig.~\ref{fig:gw} shows the accompanying GW signal extracted from the entire star with the quadrupole formula via QF1. Note that the emission arising from the prompt convection early postbounce ($\sim0.1$\,s, see e.g., \cite{marek09}) is rather weak. This is likely because the simulation first runs in 1D spherical symmetry and switches to 2D axisymmetry at $\sim$10\,ms after bounce. In this manner, the perturbation is absent for seeding the prompt convection.

\subsection{Spatial decomposition of the GW origin} \label{ssec:gw}

\begin{figure*}
    \centering
    \includegraphics[width=0.97\textwidth]{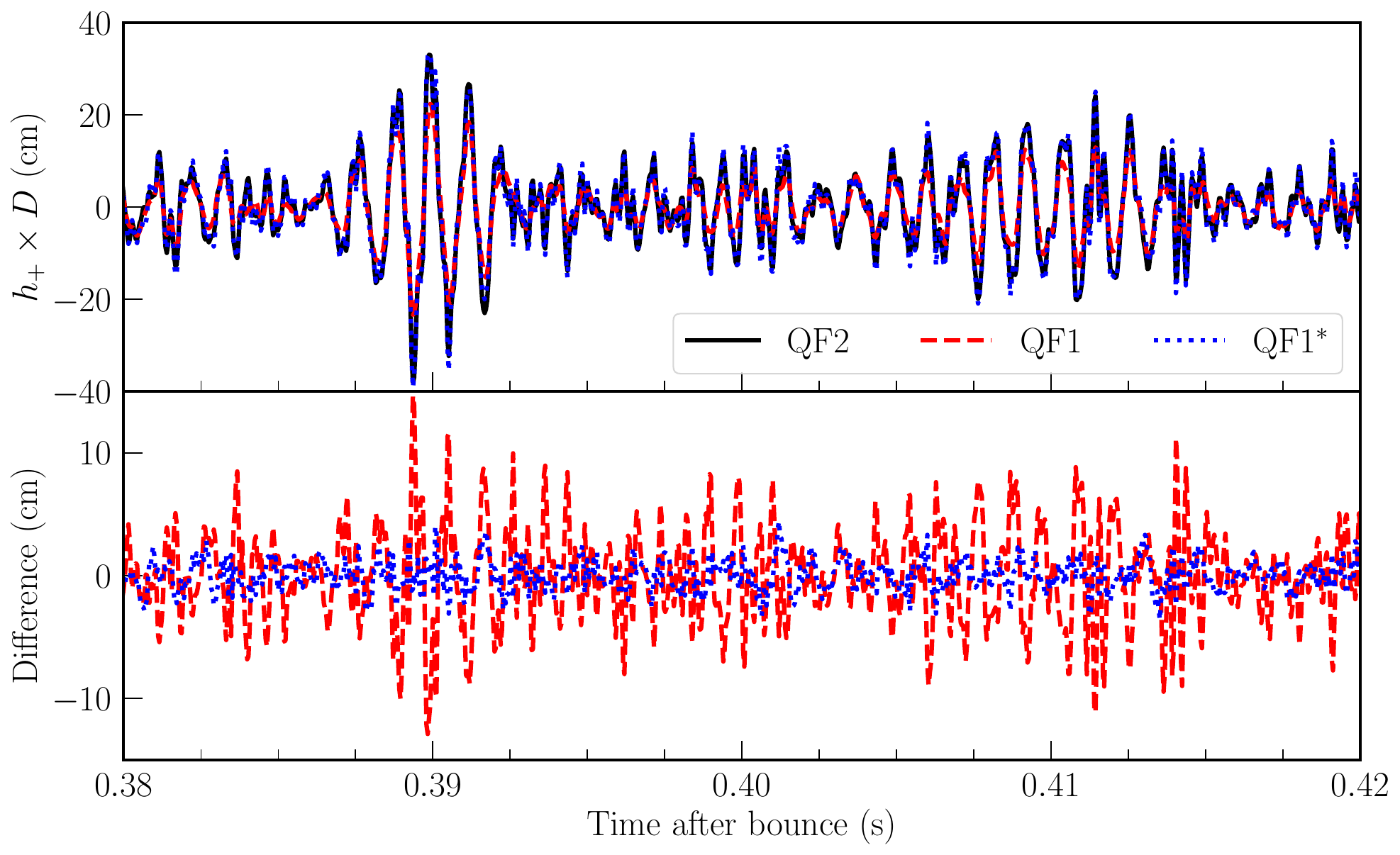}
    \caption{Comparison of the time-domain gravitational-wave waveforms extracted by the methods QF2, QF1 and QF1$^*$ for the layer in-between 20\,km and 50\,km. The lower panel shows the differences in strain amplitudes between QF2 and QF1 (red dashed) or QF1$^*$ (blue dotted).}
    \label{fig:hp_part}
\end{figure*}

\begin{figure}
    \centering
    \includegraphics[width=0.47\textwidth]{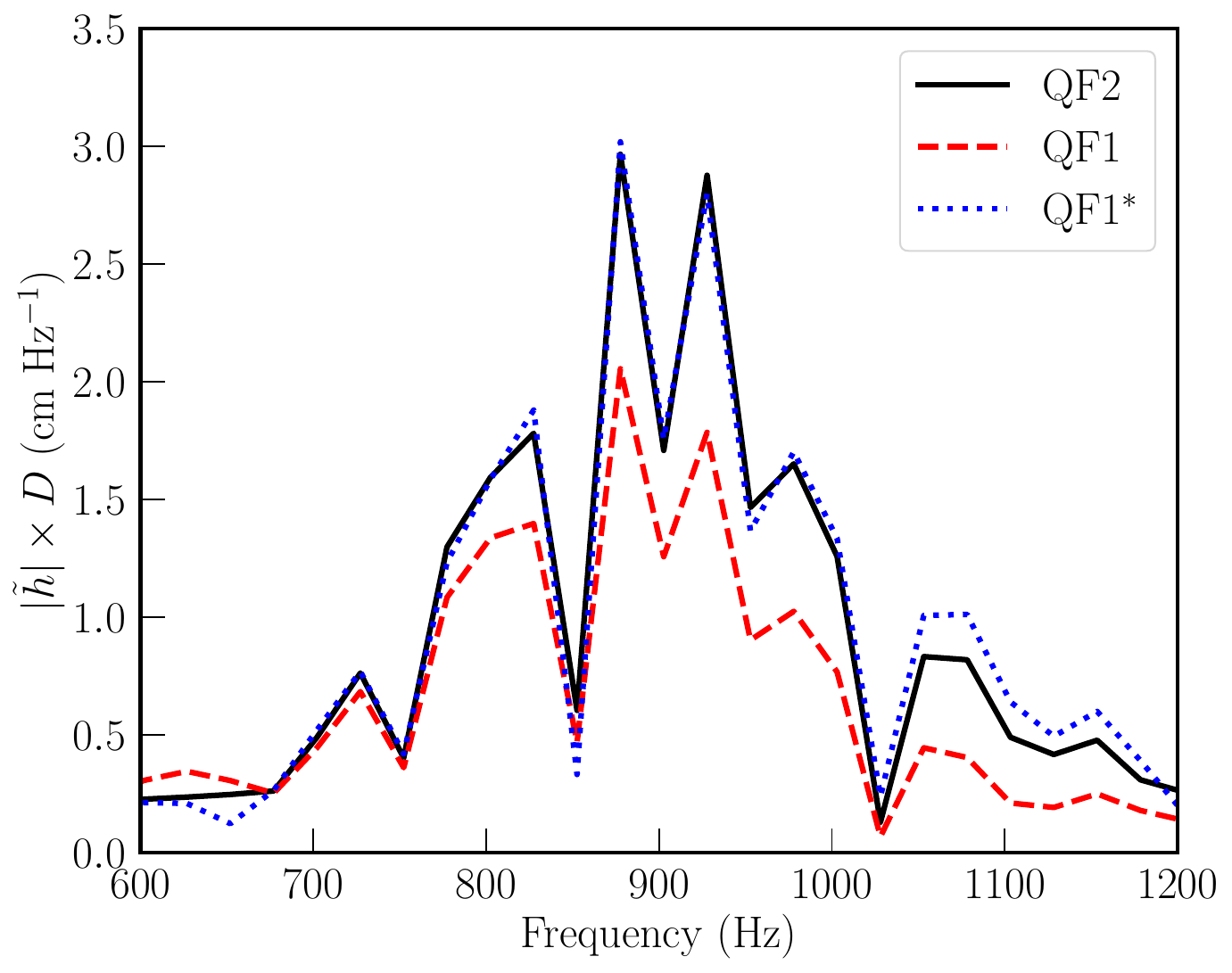}
    \caption{Fourier spectra of the gravitational-wave signals extracted by the methods QF2, QF1 and QF1$^*$ for the layer in-between 20\,km and 50\,km as shown in Fig.~\ref{fig:hp_part}. The time-domain signal is convolved with a Hann window during 0.38-0.42\,s postbounce before the Fourier transform. }
    \label{fig:hfp_part}
\end{figure}

This section presents a few numerical examples of the different methods, i.e. QF2, QF1, and QF1$^*$ (see \S\,\ref{sec:form}), for extracting GW strain amplitudes from simulations. I discuss the differences in the strain amplitude extracted by the three methods when considering a layer inside the star, and Appendix~\ref{app:hstar} demonstrates the consistency of QF2 and QF1 for the case of the entire star. Fig.~\ref{fig:hp_part} compares the corresponding time-domain GW waveforms for the layer in-between 20\,km and 50\,km in the upper panel, and the differences in strain amplitudes between QF2 and QF1 (red dashed) or QF1$^*$ (blue dotted) in the lower panel. It is quite clear that QF2 and QF1$^*$ agree well with differences below 2\,cm in $h_+\times D$, while QF1 significantly differs from them due to the ignorance of the surface term in Eq.~(\ref{eq:dIdt_layer}). The surface term can be as large as $\sim$10\,cm in $h_+\times D$, greater than about 25\% of the total amplitude. Fig.~\ref{fig:hfp_part} complements this comparison with the frequency-domain spectra of the time-domain signals convolved with a Hann window. The peak amplitude in $|\tilde{h}|\times D$ differs by about 1/3 between QF1 and QF2 (or equivalently QF1$^*$).

\begin{figure*}
    \centering
    \includegraphics[width=0.47\textwidth]{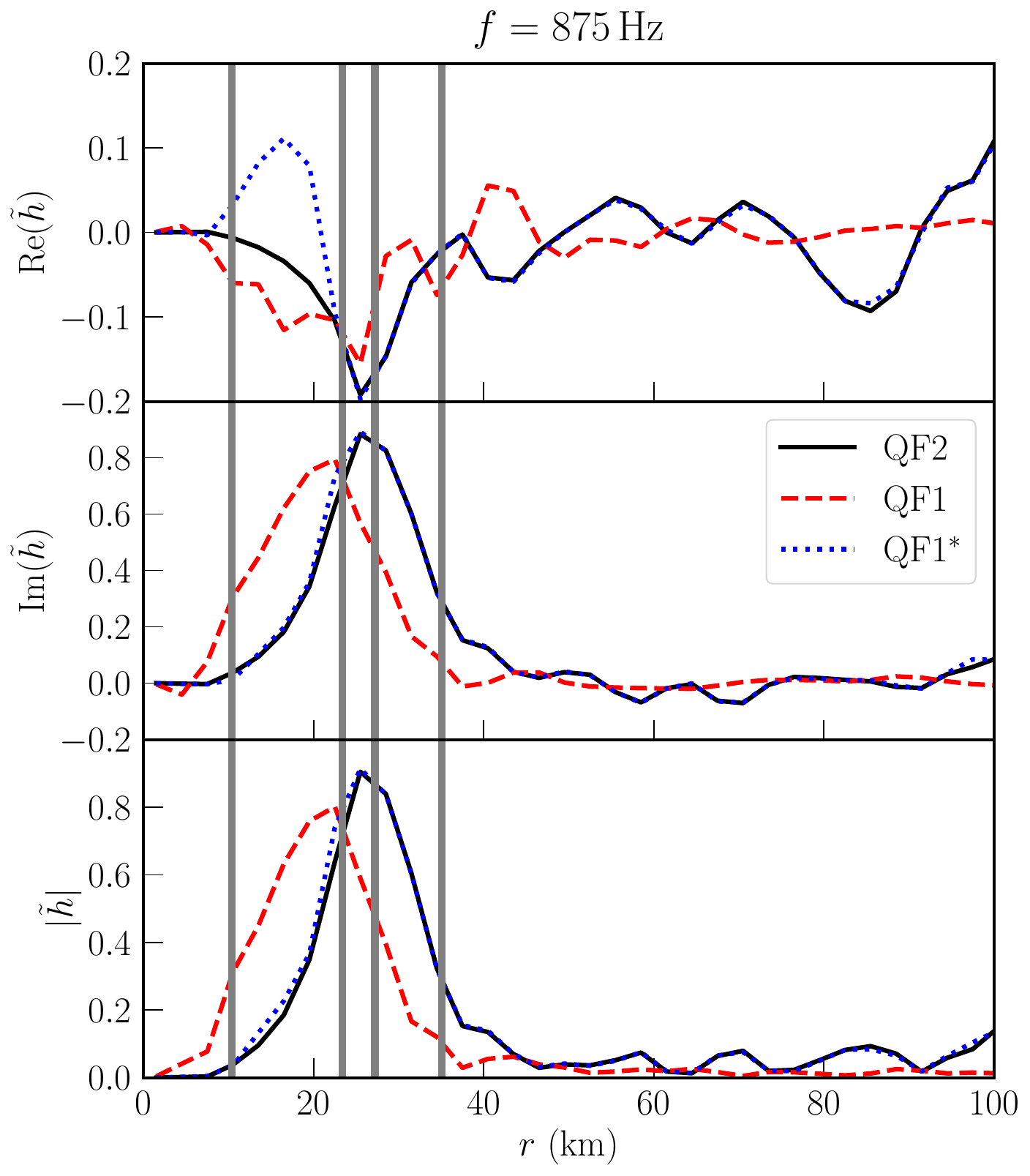}
    \includegraphics[width=0.47\textwidth]{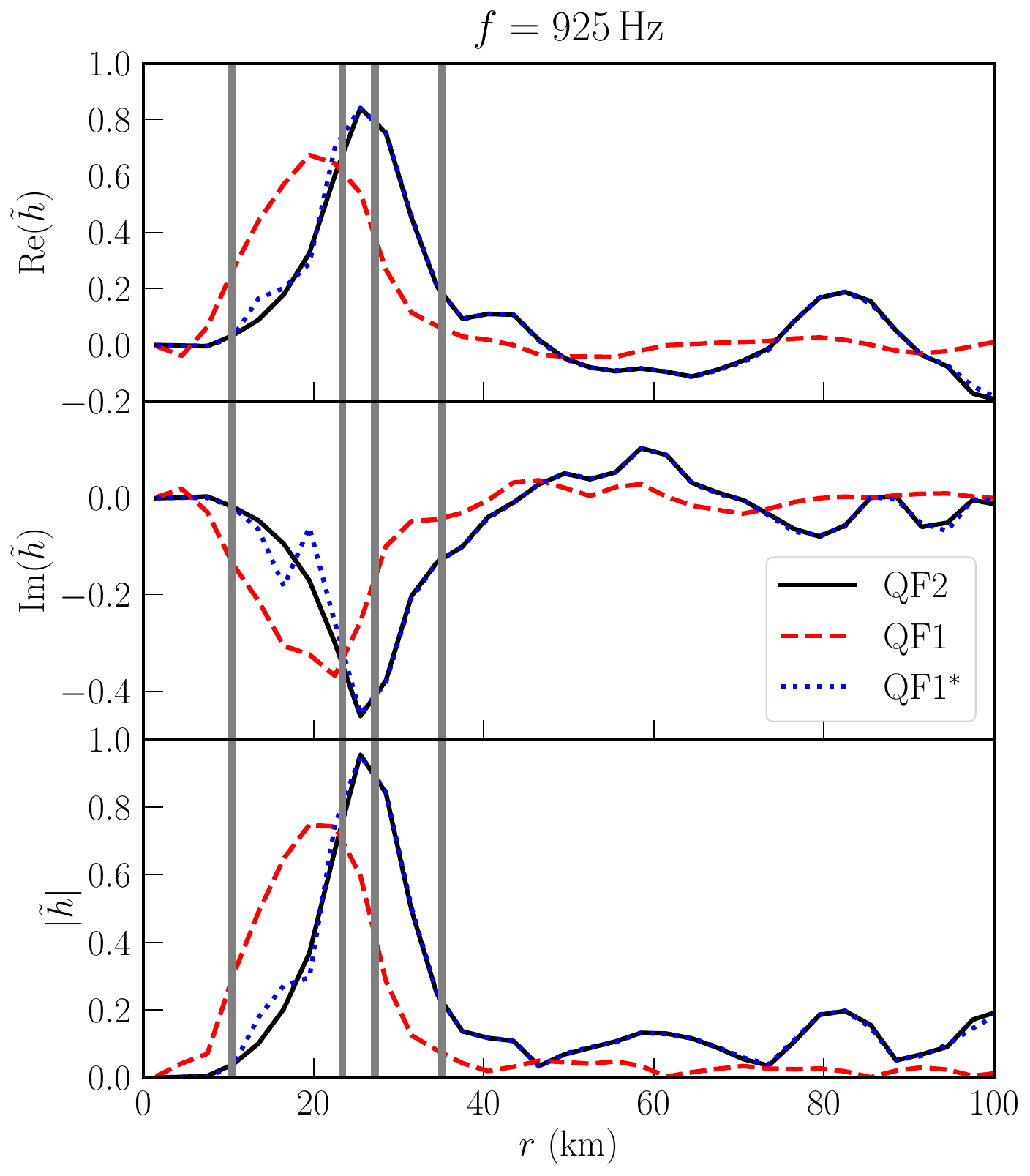}
    \caption{Comparison of the radial profiles of the complex $\tilde{h}$ and its modulus in Fig.~\ref{fig:hfp_part} at the peak frequencies $\sim875\,$Hz (left panel) and $\sim925\,$Hz (right panel) among signals extracted by the methods QF2, QF1 and QF1$^*$. I use successive layers with a thickness of 2\,km from the centre to 100\,km. From left to right, the gray vertical lines mark the lower and upper boundaries of the convective region, the upper boundary of the overshooting region, and the surface of the protoneutron star, respectively (also see Fig.~\ref{fig:sim}). The width of the vertical lines reflects the shifts in layer boundaries of $\sim0.6\,$km during 0.38-0.42\,s postbounce, cf. Fig.~\ref{fig:sim}.}
    \label{fig:hr}
\end{figure*}

Fig.~\ref{fig:hr} illustrates an additional numerical example that decomposes the peak GW emission into the contribution of successive layers with an equal thickness of 2\,km. Note that it shows the complex $\tilde{h}$ and its modulus as the fast Fourier transform produces complex spectra. Again, QF2 and QF1$^*$ show an excellent match and deviate from QF1 for the major component (imaginary part for 875\,Hz and real part for 925\,Hz) and the modulus of $\tilde{h}$. Note that the minor component shows some discrepancies for the inner core, though. The gray vertical lines in Fig.~\ref{fig:hr} mark the locations of the lower and upper boundaries of the convective region, the upper boundary of the overshooting region, and the surface of the PNS, from left to right. The width of these vertical lines reflects the shifts in layer boundaries of $\sim0.6\,$km during the period of 0.38-0.42\,s postbounce, see Fig.~\ref{fig:sim}. One can tentatively observe how the improper approach (QF1) mistakenly attributes the GW emission to different layers.  Notably, QF1 shifts the maximum contribution inwards, to the PNS convective region rather than the overshooting region as predicted by the correct methods, i.e. QF2 and QF1$^*$.

\subsection{Conventional analysis of the spatial contribution} \label{ssec:analysis}

\begin{figure*}
    \centering
    \includegraphics[width=0.97\textwidth]{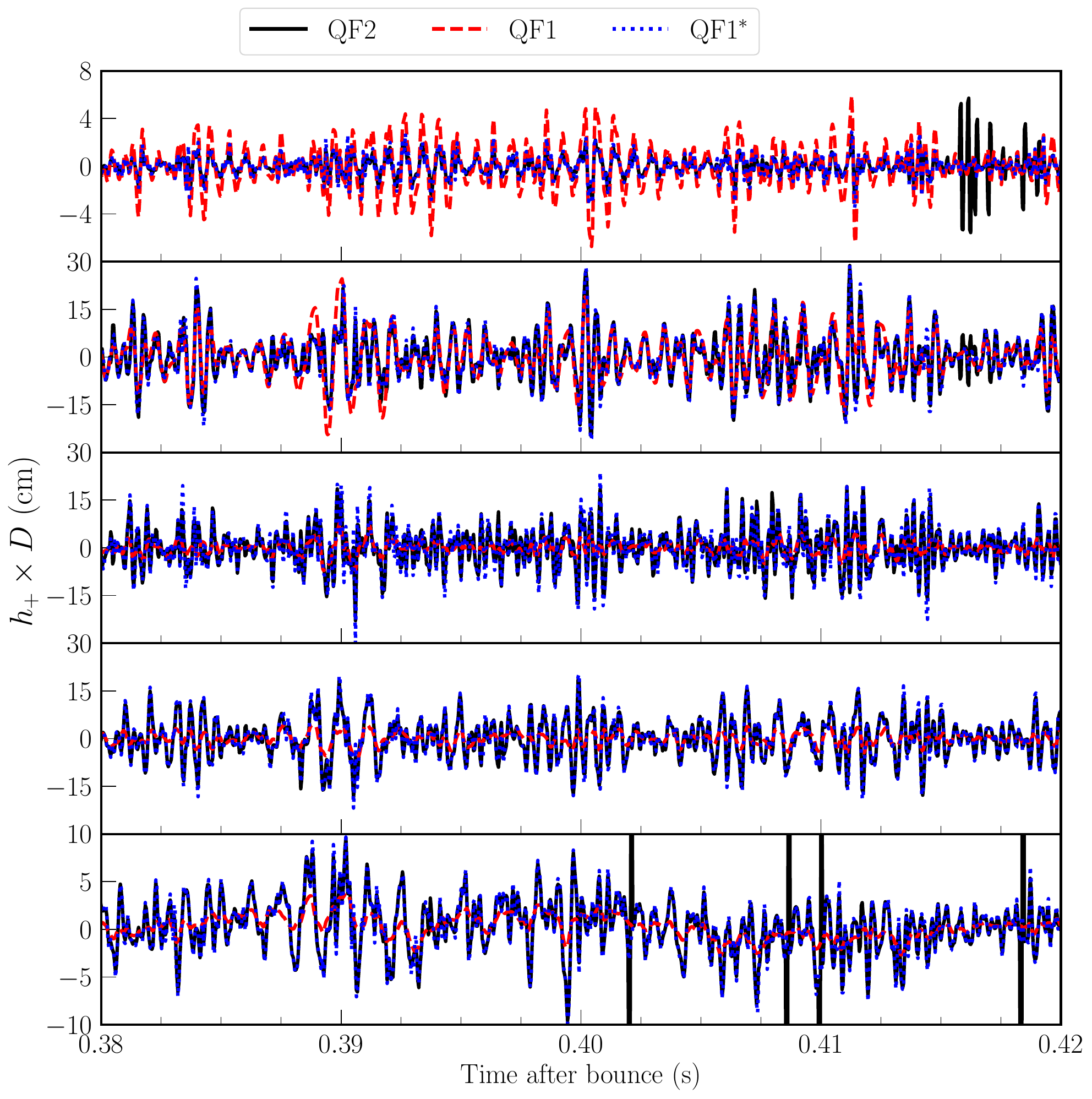}
    \caption{Comparison of the gravitational-wave waveforms contributed by different layers extracted by three methods. From top to bottom, the panels correspond to the stable core, the convective region, the overshooting region, the stable surface, and the overburden of the protoneutron star (see Fig.~\ref{fig:sim} for the division).  }
    \label{fig:ht_region}
\end{figure*}

\begin{figure}
    \centering
    \includegraphics[width=0.47\textwidth]{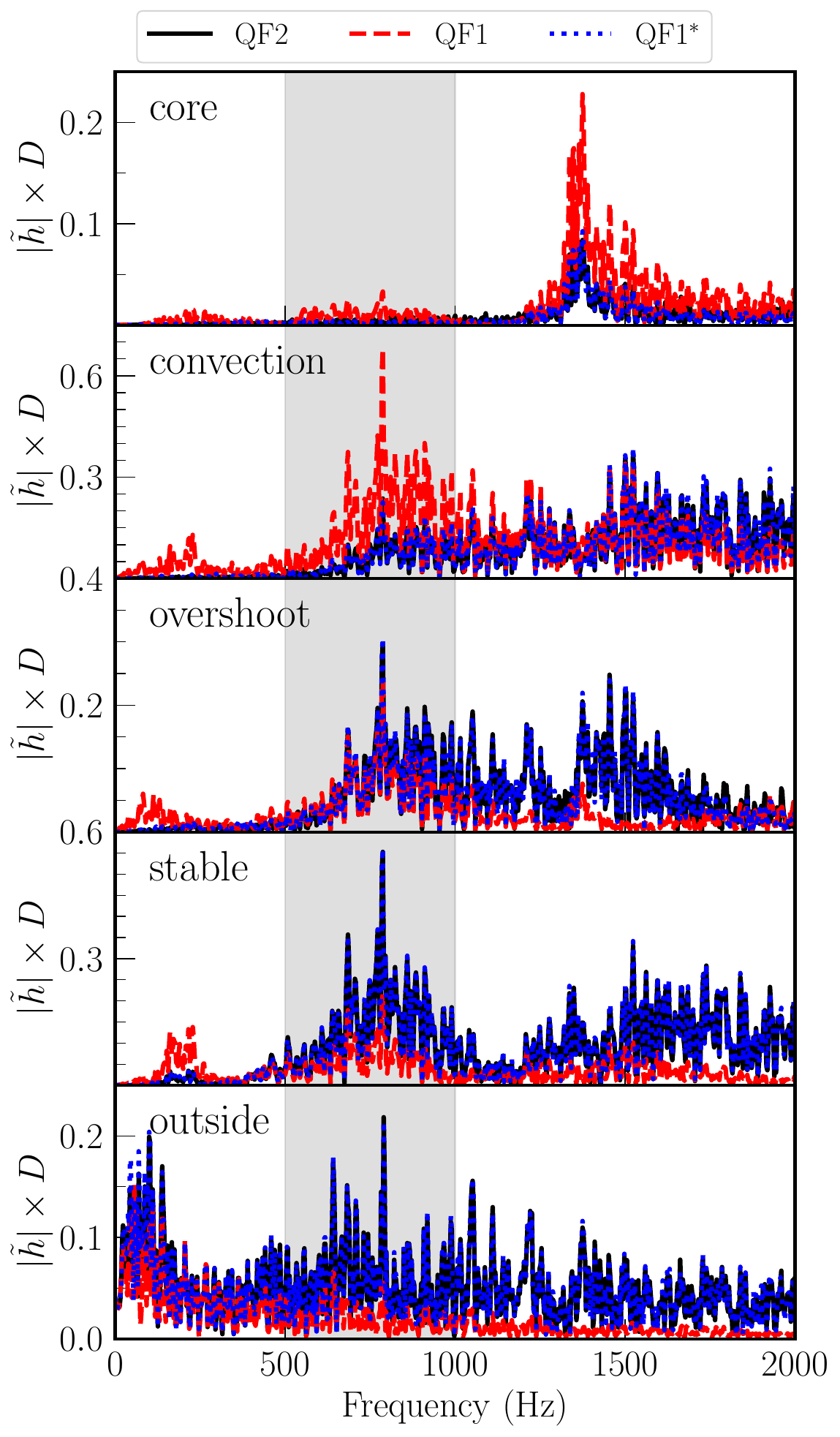}
    \caption{Fourier spectra of the gravitational-wave signals contributed by different layers extracted by three methods. The order of the panels is the same as that of Fig.~\ref{fig:ht_region}. The gray-shaded area indicates the peak emission frequency range (500-1000\,Hz, see the right panel of Fig.~\ref{fig:gw}).}
    \label{fig:hf_region}
\end{figure}

Several previous studies \cite{andresen17,mezzacappa20,mezzacappa23} analyzed the spatially dependent contribution of layers divided by the convective stability. Here I follow the approach of Andresen et al. \cite{andresen17} and divide the star into five regions, the stable PNS core, the PNS convective region, the overshooting region, the PNS stable surface, and the part outside the PNS, with the boundaries shown in the left panel of Fig.~\ref{fig:sim}. It is ambiguous to define the PNS and I use the common definition of a density threshold $10^{11}$\,g\,cm$^{-3}$. After the division, the GW strain is extracted for these five layers by the three methods, with the time-domain waveforms and Fourier spectra during 0.38-0.42\,s postbounce shown in Fig.~\ref{fig:ht_region} and Fig.~\ref{fig:hf_region}, respectively. Both figures show agreement between the results obtained by the proper methods, QF2 and QF1$^*$, while they deviate from that by the improper method QF1. In particular, the QF1 method favors the contribution from the PNS convective layer (the second row) for the peak GW emission in 500-1000\,Hz (gray-shaded area in Fig.~\ref{fig:hf_region}), while the other proper methods favor the contribution from the PNS stable surface layer (the fourth row).

\begin{figure*}
    \centering
    \includegraphics[width=0.97\textwidth]{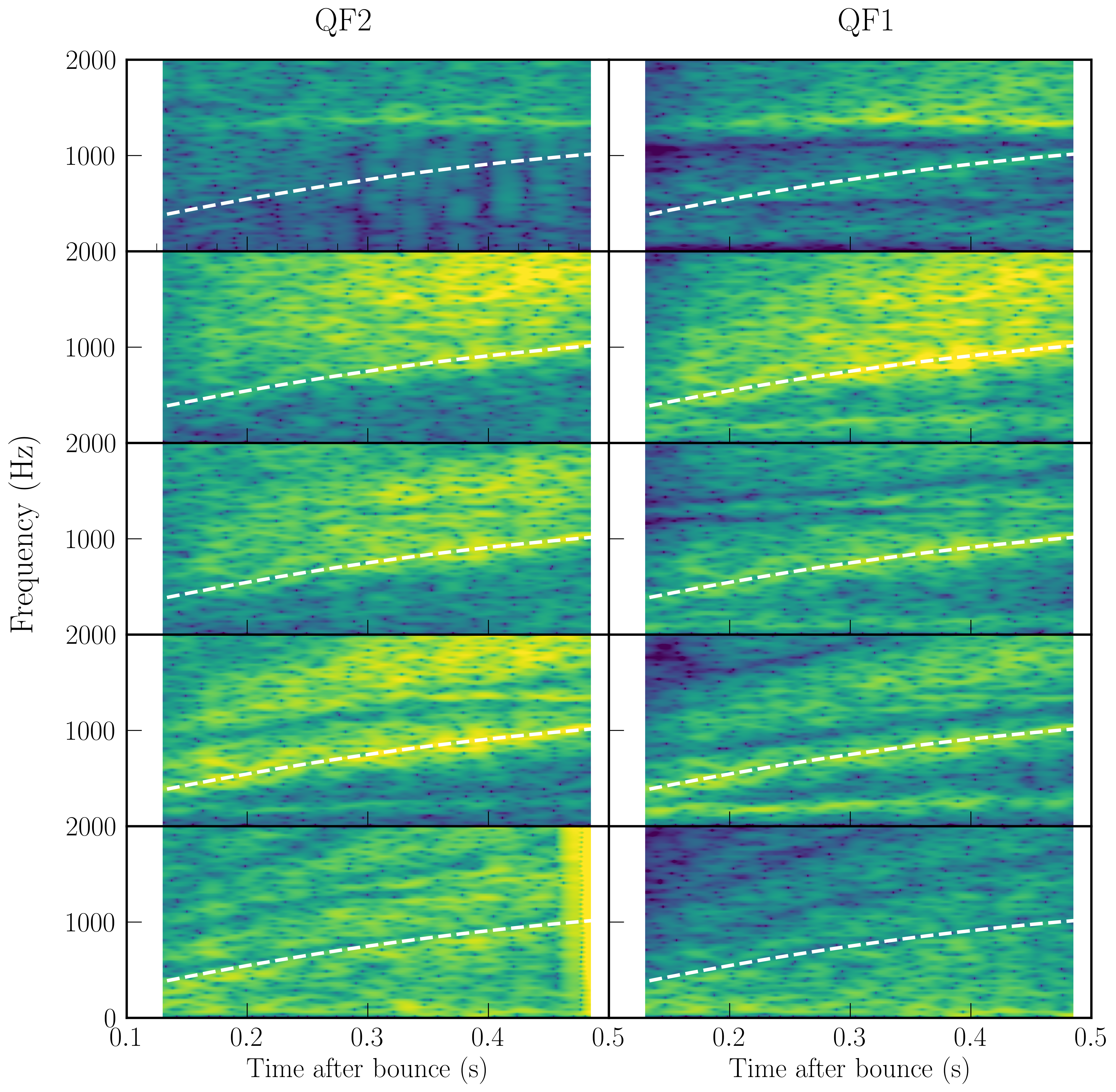}
    \caption{Spectrogram of the gravitational-wave signals contributed by different layers extracted by two methods QF2 and QF1. The vertical order of the panels is the same as that of Fig.~\ref{fig:ht_region}. The white dashed line in each panel marks the evolution of the peak emission frequency with a quadratic polynomial (Eq.~(\ref{eq:fp})). Again, I use the short-time Fourier transform with a moving temporal window of 40 ms by the Python function ‘matplotlib.pyplot.specgram'. It does not include any normalization so the color can be compared directly.}
    \label{fig:hft_region}
\end{figure*}

Fig.~\ref{fig:hft_region} further shows the GW spectrograms contributed by these five layers with the methods QF2 (the left column) and QF1 (the right column). I omit QF1$^*$ which agrees with QF2. The white-dashed line in each panel indicates the time evolution of the peak GW frequency, which is fitted with a quadratic polynomial during 0.1-0.5\,s postbounce as follows:
\begin{equation} \label{eq:fp}
    f_{\rm peak} = -2079 t_{\rm pb}^2 +3069t_{\rm pb}+15,
\end{equation}
where $t_{\rm pb}$ is the postbounce time. The conclusion is similar to Fig.~\ref{fig:hf_region} that the QF1 method favors the contribution from the PNS convective layer for the peak GW emission, while the other proper methods favor the contribution from the PNS stable surface layer. Although dimensionality can be a source of difference, the conclusion drawn from the improper method QF1 here favors the convective and overshooting regions but may overlook that from the PNS stable layer. The decomposed frequency spectra can be a better representation of the discrepancy among different methods than the time-domain waveform, which is stochastic and noisy. This is a potential caveat of their results which need confirmation in future studies.

\subsection{The global emission picture} \label{ssec:global}

\begin{figure*}
    \centering
    \includegraphics[width=0.975\textwidth]{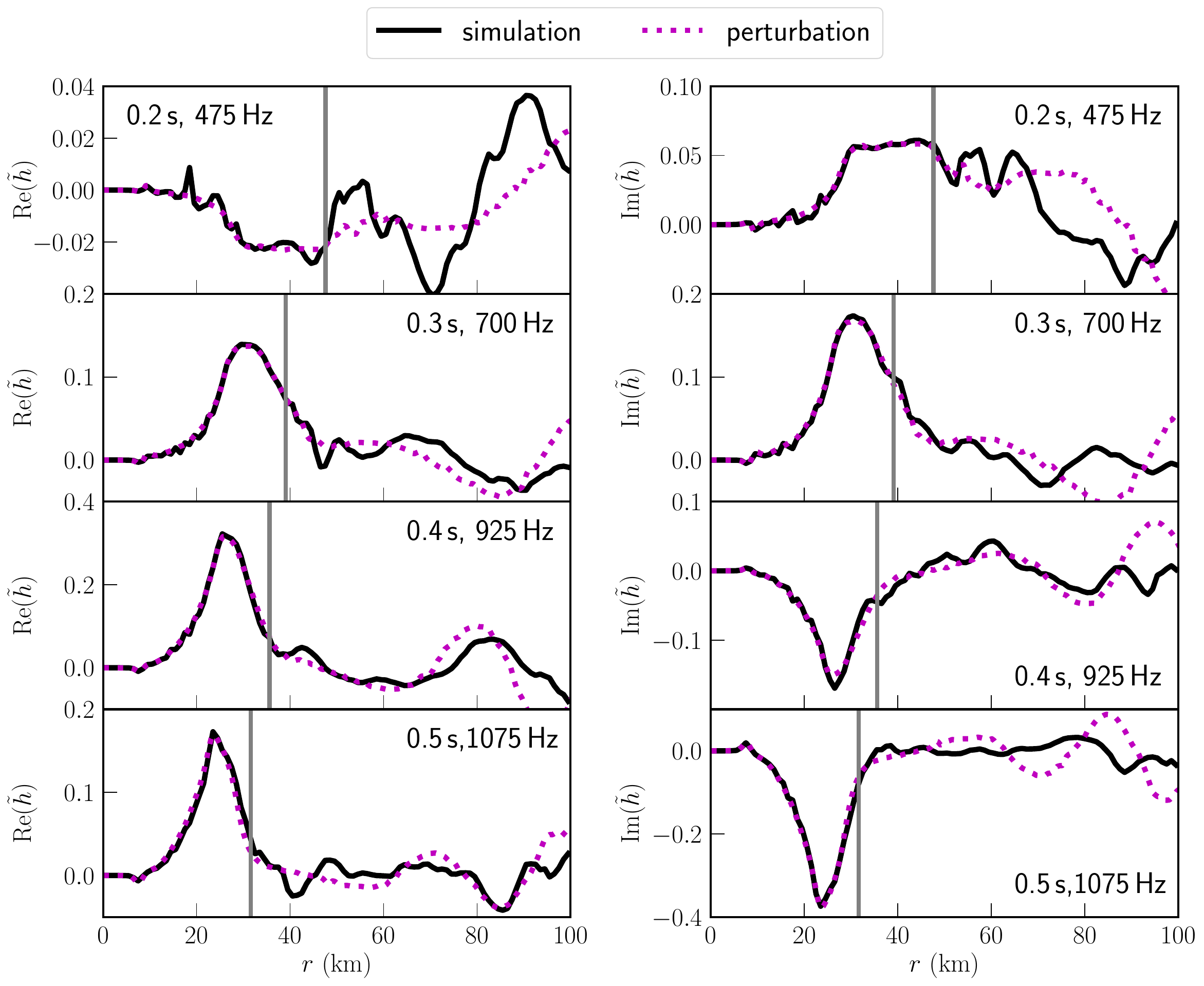}
    \caption{Matching radial profiles of gravitational-wave (GW) emission between results of the simulation and perturbative analysis. The black lines (simulation) correspond to peak GW frequencies and the magenta dotted lines (perturbation) share the same frequency. The black lines are obtained by using the method QF2 which agrees with QF1$^*$} The left and right panels show the real and imaginary parts of the Fourier components, respectively. From top to bottom, the panels show results during 0.2\,s to 0.5\,s postbounce with a 40\,ms Hann window. The gray vertical line in each panel marks the surface of the protoneutron star.
    \label{fig:global}
\end{figure*}

Zha et al. (2024) \cite{zha24} found that the radial profiles of GW emission match well at \emph{any} postbounce time and frequency between simulations and perturbative analyses in the PNS interior. Their results suggest a global emission picture for GWs from PNSs born in stellar collapses, i.e. GW emission arises from the oscillations of the PNS as a whole. This conjecture has been inferred by the coincidence between GW peak frequencies with eigenmodes acquired from perturbative analyses based on quasi-static PNS structure \cite{morozova18,torres18,radice19,sotani17}. This section replicates the analysis by matching the radial profiles of the peak GW emission for the simulation conducted in this work with the perturbative results, during the entire period with significant GW emission (0.2 to 0.5\,s postbounce).  As shown below, the correlation among GW amplitudes at different radii is determined by the PNS structure and the radial profile obtained by QF2 or QF1$^*$, but not QF1 coincides with that from a perturbative oscillation mode multiplied by a frequency-dependent constant. 

I use the tool for perturbative analyses developed in Zha et al. \cite{zha24}. In short, it solves the Newtonian perturbative equations \cite{christensen1991solar} with the PNS background model using the Runge-Kutta integration. Westernacher-Schneider \citep{ryan2020} has shown that the underlying equations are consistent with hydrodynamic equations in the pseudo-Newtonian framework.

From top to bottom, the panels in Fig.~\ref{fig:global} show the results at 4 time points, 0.2, 0.3, 0.4, and 0.5\,s postbounce with a 40\,ms Hann window. I choose the radial profiles of GW emission at the corresponding frequency for the comparison between simulation and perturbative analyses. The simulation results are obtained by using the method QF2 that agrees with QF1$^*$ (cf. Fig.~\ref{fig:ht_region}). Perturbative functions are multiplied with constants to match the simulation results in the PNS, marked by the gray vertical lines. Fig.~\ref{fig:global} re-emphasizes the global emission picture of GW emission in stellar collapses presented by Zha et al. \cite{zha24}. 

\section{Conclusions and outlook} \label{sec:conclu}

In conclusion, I have discussed the methods for computing the spatially dependent contribution of GW signals in stellar collapse simulations. The methods are valid in the framework of quadrupole formulae within the approximations of slow motion and weak fields for pseudo-Newtonian simulations. I presented three methods:
\begin{enumerate}
    \item QF2 which takes the numerical differentiation of trace-free quadrupole moment twice, Eq.~\ref{eq:Izz};
    \item QF1 which reduces the numerical differentiation to only once with the help of mass conservation, Eq.~\ref{eq:dIdt};
    \item QF1$^*$ which corrects QF1 with the surface term from the partial integral, Eq.~\ref{eq:dIdt_layer}.
\end{enumerate}
An important note is that when considering the emission of a layer, QF2 and QF1$^*$ are mathematically equivalent and the proper methods. The surface term in QF1$^*$ is from direct mathematical derivation and one should not confuse it with the ambiguous determination of particular boundaries.

I demonstrate their usage with numerical experiments, a 2D axisymmetric pseudo-Newtonian simulation for the collapse and explosion of a 20~$M_\odot$ star. Indeed QF2 and QF1$^*$ agree well numerically and QF1 introduces significant bias. The improper method QF1 mistakenly favors the dominant contribution by the PNS convective layer. By comparison with the results of perturbative analyses, I emphasize again the global picture of the GW emission from PNS oscillations.

Properly decomposing the GW origin can aid the understanding of GWs from stellar collapses together with the analysis of the complex fluid motions and other diagnostics as done in \cite{andresen17,mezzacappa20,mezzacappa23}. A combined analysis with both the QF2 and QF1$^*$ methods, as conducted in this study, is optimal for mitigating potential numerical issues related to integration and differentiation. Moreover, the comparison of simulation and perturbative results similar to \S\,\ref{ssec:global} is currently absent in 3D, and it is highly appreciated if one has sufficient computational resources. 

Another important caveat for similar studies is the general relativistic nature of GW emission. It is important to recognize that the widely used quadrupole formula represents a pragmatic simplification of the underlying physics \cite{thorne80}. GWs are coupled with matter, space, and time. The localization of GW emission is not valid in principle. Nonetheless, attributing GW emission to different regions according to the quadrupole formulae is beneficial for connecting the complex hydrodynamic behavior with features of GW emission for concurrent stellar collapse simulations \cite{andresen17,mezzacappa20,mezzacappa23}. Advancing our comprehension of these effects awaits future developments on integrating general relativity in simulation models.

\section*{Data availability}
All the data and analysis scripts are publicly available for reproducing figures in this paper at Zenodo (doi: 10.5281/zenodo.13743902).

\section*{Software}
\textsc{FLASH} \citep{FLASH,oconnor18a}; \textsc{NuLib} \citep{gr1d};\textsc{yt} \citep{yt}; \textsc{Numpy} \citep{numpy};  \textsc{SciPy} \citep{scipy}; \textsc{Matplotlib} \citep{matplotlib}.

\begin{acknowledgments}
I thank Evan O'Connor and Sean Couch for the \texttt{FLASH} development and Evan for stimulating discussions. I also thank Dr. Zhenyu Zhu for pointing out the simplification of the quadrupole formula and the GR nature of GW emission. I gratefully acknowledge Dr. Zonglin Yi for her careful proofreading.
This work is supported by the National Natural Science Foundation of China (NSFC, Nos. 12288102, 12393811, 12090040/3), the National Key R\&D Program of China (Nos. 2021YFA1600401 and 2021YFA1600400), the International Centre of Supernovae, Yunnan Key Laboratory (No. 202302AN360001) and the Natural Science Foundation of Yunnan Province (No. 202201BC070003) and the Yunnan Fundamental Research Project (No. 202401BC070007). The authors gratefully acknowledge the “PHOENIX Supercomputing Platform” jointly operated by the Binary Population Synthesis Group and the Stellar Astrophysics Group at Yunnan Observatories, CAS.
\end{acknowledgments}

\appendix

\section{Gravitational-wave extraction for the entire system \label{app:hstar}}
\begin{figure*}
    \centering
    \includegraphics[width=0.95\textwidth]{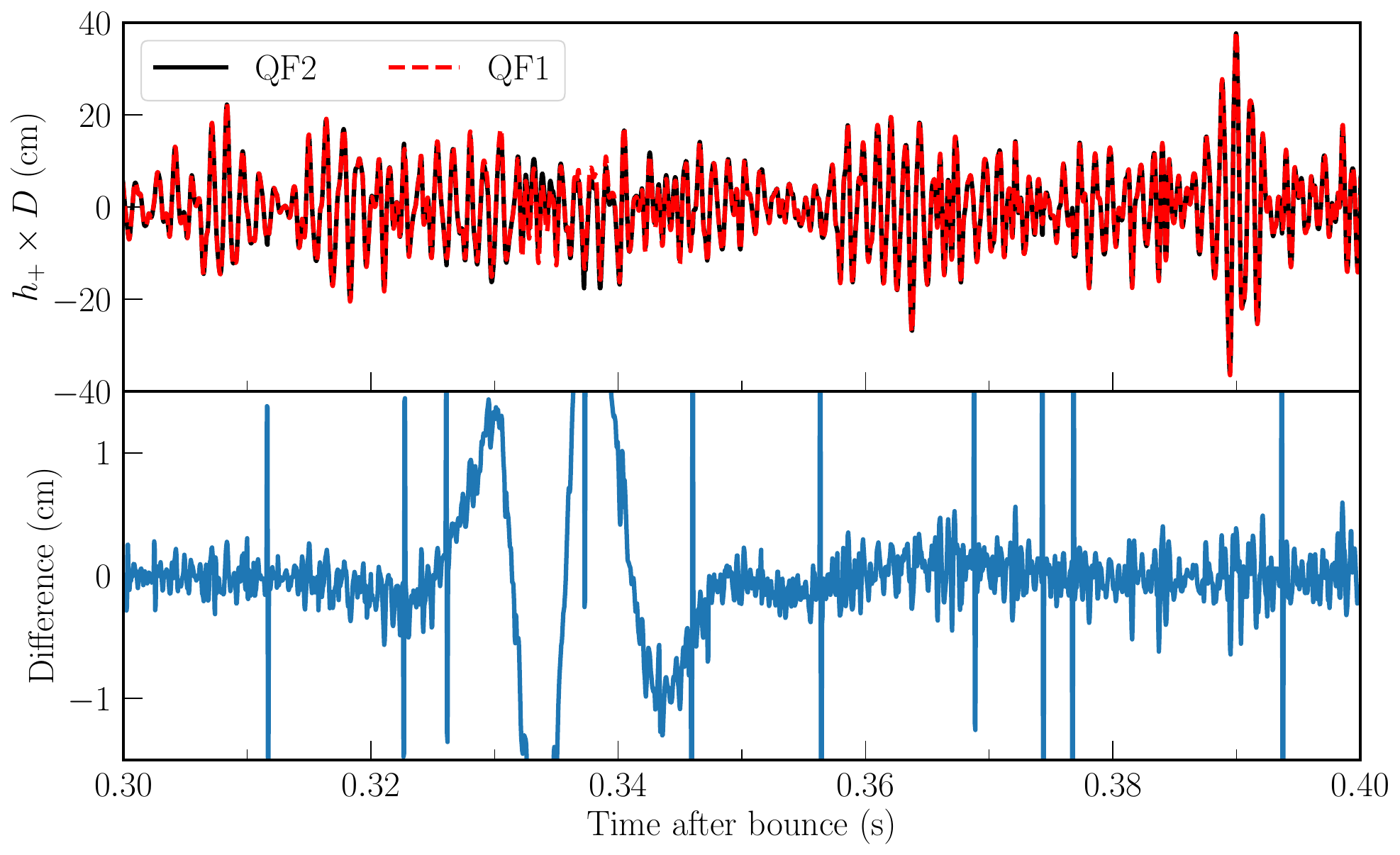}
    \caption{Comparison of the time-domain gravitational-wave waveforms extracted by the methods QF2 and QF1  (same as Fig.~\ref{fig:gw}) for the whole star. The lower panel shows the differences in strain amplitudes. The spikes in the lower panel originate from numerical finite differences. The significant deviation of $\sim1$\,cm during 0.32 to 0.35\,s is due to the outer regions, far from the protoneutron star. Its frequency is relatively low, $\sim100$\,Hz, so it does not affect the peak emission. }
    \label{fig:hp_star}
\end{figure*}

\begin{figure}
    \centering
    \includegraphics[width=0.47\textwidth]{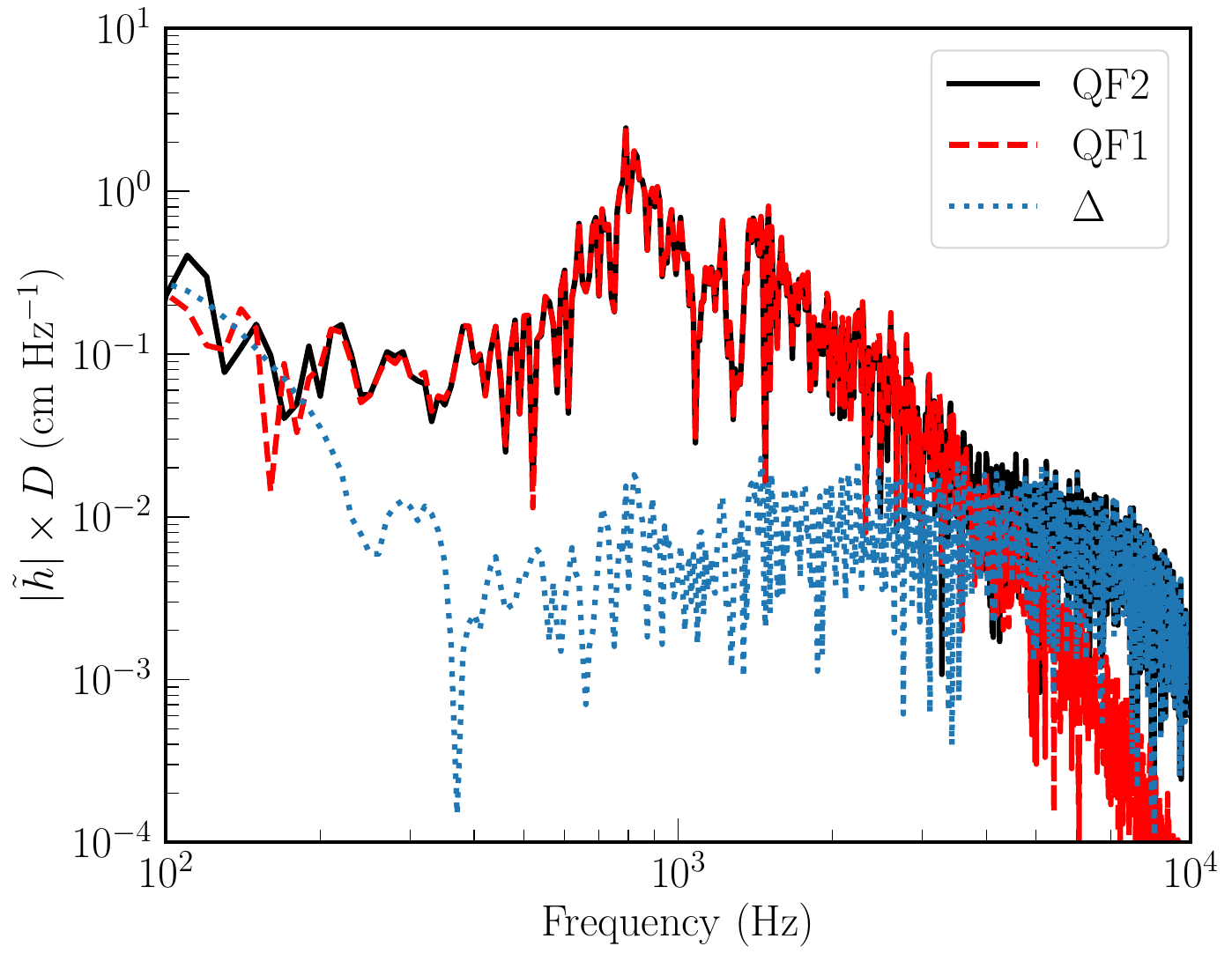}
    \caption{Fourier spectra of the gravitational-wave signals extracted by the methods QF2 and QF1 as shown in Fig.~\ref{fig:hp_star}. The time-domain signal is convolved with a Hann window during 0.3-0.4\,s postbounce before the Fourier transform.  $\Delta$ is the Fourier spectrum of the differences in the strain amplitude, i.e. the lower panel in Fig.~\ref{fig:hp_star}, not the difference between their Fourier spectra. }
    \label{fig:hfp_star}
\end{figure}

It is well known that the QF1 and QF2 methods are consistent for the gravitational-wave emission of the whole star (e.g., \cite{finn90}). For completeness and to demonstrate the numerical quality, I checked this consistency based on the simulation run in this work. Fig.~\ref{fig:hp_star} shows the time-domain GW waveforms extracted with QF2 and QF1 in the upper panel, and their differences in the lower panel. It only includes the interval of 0.3-0.4\,s postbounce for a clear presentation. As QF2 and QF1 are mathematically equivalent for the whole star, the two waveforms agree well with differences smaller than $\sim0.5$\,cm in $h_+\times D$, about $2.5\%$ of the total amplitude. In the lower panel, spikes appear occasionally due to numerical finite differences. The significant deviation of $\sim1$\,cm during 0.32 to 0.35\,s is due to the outer regions, far from the PNS. Its frequency is relatively low, $\sim100$\,Hz, so it does not affect the peak emission which is always above $100$\,Hz. Fig.~\ref{fig:hfp_star} complements this comparison with the frequency-domain spectra of the time-domain signals convolved with a Hann window. Note that $\Delta$ is the Fourier spectrum of the differences in the strain amplitude, i.e. the lower panel in Fig.~\ref{fig:hp_star}, not the difference between the other two Fourier spectra. The difference in spectra between QF2 and QF1 is mainly significant at high frequencies, $>\sim3000$\,Hz. This is expected because taking an additional numerical finite difference introduces more high-frequency noises. 

\bibliography{apssamp}% Produces the bibliography via BibTeX.

\end{document}